\documentclass{moriond}
\usepackage{amssymb}
\usepackage{amscd}
\usepackage{latexsym}
\usepackage{slashed}
\usepackage{color}
\usepackage{graphicx}
\usepackage[normalem]{ulem}
\usepackage{color}
\definecolor{orange}{cmyk}{0,0.5,1,0}
\usepackage{verbatim}
\usepackage{rotating}
\usepackage{braket}
\usepackage{diagbox}

\def\lsim{\raise0.3ex\hbox{$\;<$\kern-0.75em\raise-1.1ex\hbox{$\sim\;$}}}
\def\gsim{\raise0.3ex\hbox{$\;>$\kern-0.75em\raise-1.1ex\hbox{$\sim\;$}}}
\usepackage{graphicx,multirow,subfigure}

\newcommand{\snu}{\tilde{\nu}}
\newcommand{\snuRI}{\snu _1 ^{{\rm R,I}}}

\def\be{\begin{equation}}
\def\ee{\end{equation}}
\def\bea{\begin{eqnarray}}
\def\eea{\end{eqnarray}}

\def\be{\begin{equation}}
\def\ee{\end{equation}}
\def\bea{\begin{eqnarray}}
\def\eea{\end{eqnarray}}

\newcommand{\Photo}{}

\begin{document}
\vspace*{-1cm}
\title{Sneutrino Dark Matter, Constraints and Perspectives}

\author{Luigi Delle Rose,}
\address{School of Physics and Astronomy, University of Southampton, Highfield, Southampton SO17 1BJ, United Kingdom. }
\author{Shaaban Khalil,}
\address{Center for Fundamental Physics, Zewail City of Science and Technology, Sheikh Zayed,12588 Giza, Egypt}
\author{\textbf{Simon J.D. King}\footnote{Speaker},}
\address{School of Physics and Astronomy, University of Southampton, Highfield, Southampton SO17 1BJ, United Kingdom. }
\author{Suchita Kulkarni,}
\address{Institut f{\"u}r Hochenergiephysik, {\"O}sterreichische Akademie der Wissenschaften, 
	Nikolsdorfer Gasse 18, 1050 Wien, Austria}
\author{Carlo Marzo,}
\address{National Institute of Chemical Physics and Biophysics, R{\"a}vala 10, 10143 Tallinn, Estonia}
\author{Stefano Moretti}
\address{School of Physics and Astronomy, University of Southampton, Highfield, Southampton SO17 1BJ, United Kingdom. }
\address{Particle Physics Department, Rutherford Appleton Laboratory, Chilton, Didcot, Oxon OX11 0QX, United Kingdom}
\author{and Cem S. Un}
\address{Department of Physics, Uluda\~{g} University, TR16059 Bursa, Turkey}

\maketitle
\abstracts{The prevalent Dark Matter (DM) candidate of the $(B-L)$ Supersymmetric Standard Model (BLSSM) is the Right Handed (RH) sneutrino. In this work we assess the ability of ground and space based experiments to establish the nature of this particle, through indirect and collider detection. }

\section{Introduction}
Low scale Supersymmetry (SUSY) offers a solution to many of the major flaws of the Standard Model (SM), such as the  hierarchy problem, the absence of gauge coupling unification and several others. In addition, SUSY provides a candidate for DM, when $R-$parity conservation is imposed, which in turn requires the Lightest Supersymmetric Particle (LSP) to be stable. In the Minimally Supersymmetric Standard Model (MSSM), a possible DM candidate is the LSP neutralino, which can be bino, wino or higgsino-like. In the Constrained MSSM (CMSSM) case, the most prevalent candidate is a bino-like neutralino. In this work we adopt a similar framework for an extension of the MSSM with a gauged (\textit{B-L}) symmetry, the BLSSM. By extending the SM gauge group to $SU(3)_c \times SU(2)_L \times U(1)_Y \times U(1)_{B-L}$, one finds many more DM candidates in addition to the bino-like LSP \cite{DelleRose:2017hvy,DelleRose:2017smp,DelleRose:2017ukx}. This extension requires three RH neutrinos be added and their SUSY partners to cancel the $U(1)_{B-L}$ anomalies. One may solve an additional problem in the SM, of light non-vanishing neutrino masses, by implementing a Type-I see-saw mechanism. Unlike the usual case with very heavy RH neutrinos, we use a low (TeV)-scale see-saw, where the mass of the RH is set by the $U(1)_{B-L}$ breaking scale $\sim$ TeV. Consequently, one requires small Yukawa couplings, $Y_\nu \sim \mathcal{O}(10^{-6})$ to explain the light neutrino masses.

The superpartner of these neutrinos, the RH sneutrinos, provide another DM candidate, which can easily comply with direct detection and relic abundance requirements \cite{DelleRose:2017uas}. In this work, we investigate the feasibility to detect such a DM candidate. We begin by studying the annihilation properties of the sneutrino in section 2. Then we consider indirect detection in section 3. Finally we move to the collider signatures in section 4 and then conclude.

\section{Annihilation Properties}
Due to a lepton number violating operator, the RH sneutrino and anti-sneutrino are no longer the mass eigenstates. The LSP can either be the CP even sneutrino, $\tilde{\nu}_1^{\rm Re}$, or a CP odd sneutrino, $\tilde{\nu}_1^{\rm Im}$. This has important consequences for phenomenology, and forbids the annihilation of two LSPs into $Z'$. The relic abundance of the sneutrino DM is a direct consequence of the strength of the DM annihilations, and the type of allowed interaction will determine what collider signatures this DM candidate may provide. The main interactions which contribute to the annihilations of the sneutrino DM are given by  four-point interaction $\left( \tilde{\nu}^{({\rm R,I})}_{{1}} \tilde{\nu}^{({\rm R,I})}_{{1}} \rightarrow h_i h_j \right)$ and processes mediated by the CP-even Higgs sector $\left(\tilde{\nu}^{({\rm R,I})}_{{1}} \tilde{\nu}^{({\rm R,I})}_{{1}} \rightarrow  h_i \rightarrow h_i h_j~{\rm or} ~W^+ W^- \right) $. 

When kinematically allowed, the dominant annihilation channel for the sneutrino LSP is into the lightest (\textit{B-L}) Higgs state, ie $\tilde{\nu}^{({\rm R,I})}_{{1}} \tilde{\nu}^{({\rm R,I})}_{{1}} \rightarrow h_2 h_2$. When this is disallowed $(M_{\tilde{\nu}}<M_{h_2})$, the dominant channel is to decay to a $W^+ W^-$ pair. The existence of this channel is crucial to indirect detection, as we will discuss in the next section. The value of these annihilation cross sections determine the relic abundance. In this work we consider a standard cosmological scenario, where the DM particles were in thermal equilibrium with the SM ones  in the early Universe and decoupled when the temperature fell below their relativistic energy. The relic density of our sneutrino species is written as \cite{Lee:2007mt}:
\begin{equation}
\Omega h^2 _{\snu _1 ^{{\rm R,I}}} = \frac{2.1 \times 10^{-27} \rm{cm}^3 \rm{s}^{-1}}{\braket{\sigma_{\snu _1 ^{{\rm R,I}}}^{\rm{ann}}v}} \left( \frac{x_F}{20}\right) \left(\frac{100}{g_* (T_F)}\right) ^{\frac{1}{2}},
\end{equation}
where $\braket{\sigma_{\snu _1 ^{{\rm R,I}}}^{\rm{ann}}v}$ is a thermal average for the total cross section of annihilation to SM objects multiplied by the relative sneutrino velocity, $T_F$ is the freeze out temperature, $x_F \equiv m_{\snuRI}/T_F \simeq \mathcal{O}(20)$ and $g_*(T_F) \simeq \mathcal{O}(100)$ is the number of degrees of freedom at freeze-out.

\begin{figure}[h]
	\vspace{-0.3cm}
	\centering
	\includegraphics[width=10cm, height=6cm]{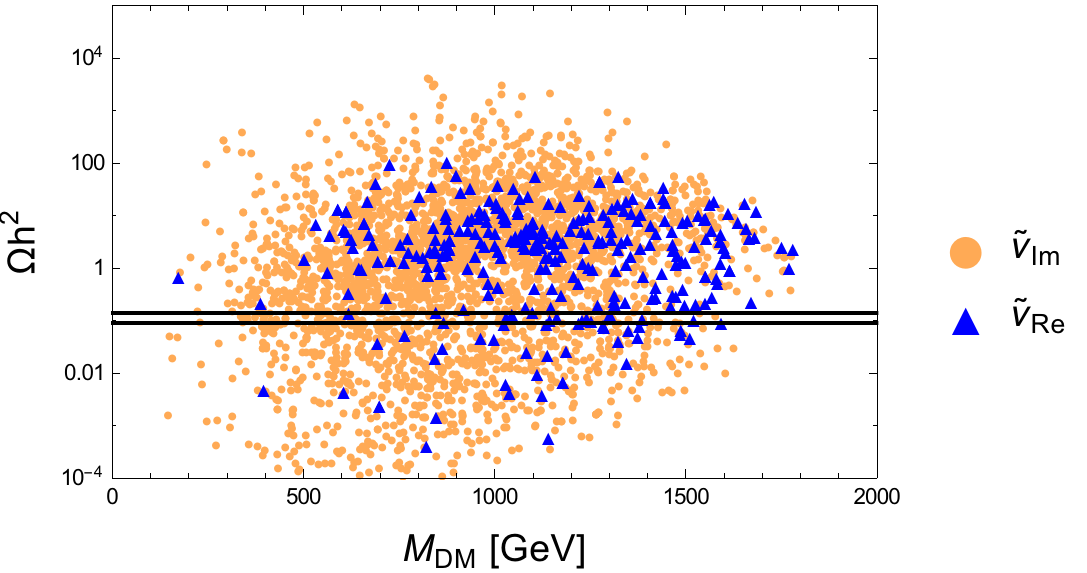}
	\caption{Relic density of CP-even and CP-odd sneutrinos versus their mass in GeV, where horizontal lines correspond to the Planck limits for the relic abundance.}
	\label{fig:Sneutrino_Relic}
	\vspace{-0.0cm}
\end{figure}

Fig. \ref{fig:Sneutrino_Relic} shows the thermal relic abundance for sneutrinos. This has been computed by micrOMEGAs  \cite{Belanger:2006is,Belanger:2013oya} and one can see that both CP-even and CP-odd candidates are allowed by current limits of $0.09 < \Omega h^2 < 0.14$, which is the $2\sigma$ allowed region by the Planck collaboration \cite{Ade:2015xua}.

\section{Indirect Searches}
\begin{figure}[h]
	\centering
	\includegraphics[width=12cm, height=7.5cm]{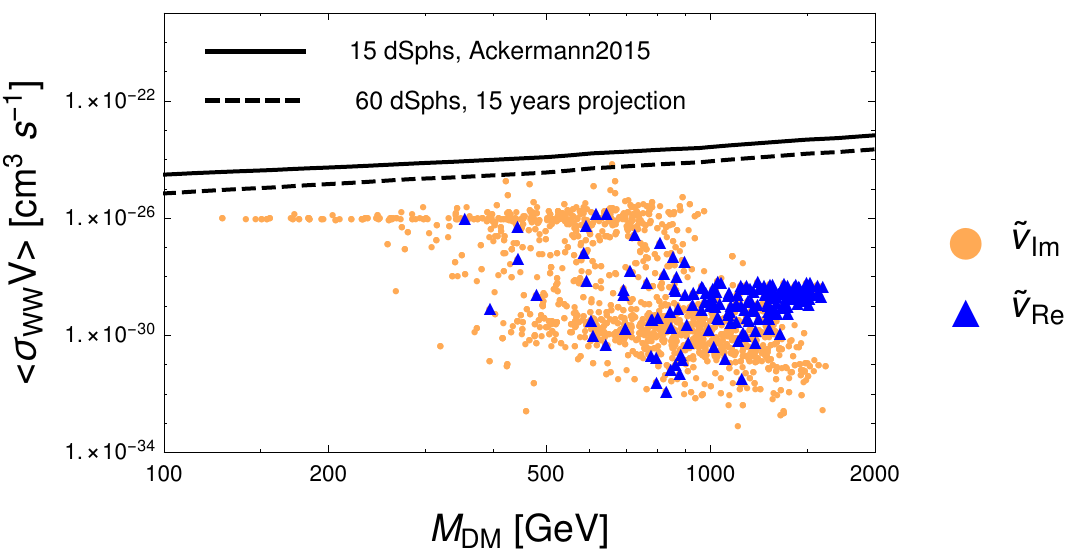}
	\caption{Thermal cross section for DM DM $\to W^+W^-$ annihilation as predicted by theory as a function of the DM mass, for  
		CP-even (blue) and CP-odd (orange) sneutrinos. Also shown are the Fermi-LAT limit from dSphs at present (solid black) and as projection for 15 years from now (dashed black).
		All points obey the relic density upper limit, for which rescaling, where necessary, has been applied.}
	\label{fig:Scan-FermiLAT}
\end{figure}

When two sneutrinos annihilate in the galactic centre, they may produce charged $W^+W^-$ pairs, which may radiate off high energy gamma rays, which can be observed by the Fermi-LAT experiment \cite{Ackermann:2015zua}. In Fig.~\ref{fig:Scan-FermiLAT}, we plot the sneutrino annihilation cross section in the $W^+W^-$ channel. We choose to split populations of CP-odd, $\snu _{Im}$ and CP-even, $\snu _{Re}$, to demonstrate that both sectors may be viably explored. The existence of a spectrum point which may be observed in the projected 15 years data sample, shows that the experiment is just now beginning to touch the relevant parameter space.
\section{Collider Signatures}
In searching for these
direct DM signals,
we have scanned over several benchmark CP-even and CP-odd sneutrino LSPs and used {MadGraph} \cite{Alwall:2014hca} for the computation of the LHC cross sections. One may search for the sneutrino LSP through typical DM searches, such as mono-jet or VBF with invisible Higgs decays, but these do not uniquely identify the model. Production via $pp \rightarrow Z' \rightarrow \snu ^{\rm I}_1 \snu ^{\rm R}_i, \snu ^{\rm R}_1 \snu ^{\rm I}_i$  would provide an interesting signature, but due to the mass limit on the $Z'$, $M_{Z'}>4$TeV, the cross section is at most $\sigma \simeq 0.025$, so a signal would not be observable in the near future. 

A promising signature of the sneutrino LSP could be done indirectly, via slepton pair production. We have found benchmark points with a cross section $\sigma \sim \mathcal O(0.1)$ fb's. In scenarios with light sleptons, one may find that the $\tilde l \rightarrow W^\pm \snu_{LSP}$ channel is the only available decay mode despite a width suppressed by the small Dirac Yukawa coupling, and this would yield a di-lepton signature.

For a different mass hierarchy, one can have $\tilde l \rightarrow \tilde \chi^0 l$ with $\tilde \chi^0 \rightarrow \nu_h \snu_{LSP}$, where $\nu_h$ is the heavy neutrino. The latter will mainly undergo $\nu_h \rightarrow W^\pm l^\mp$ or $\nu_h \rightarrow Z \nu_l$ decay, thus providing fully or semi-leptonic signatures (again, accompanied by missing transverse energy). Finally, one may observe DM signatures from the pair production of squarks, which can have large cross sections, $\sigma \sim \mathcal{O}(1)$fb's. One may find sneutrinos through a squark decay chain $\tilde t \rightarrow \tilde \chi^0 \, t$, with a large branching fraction $\sim$ 80\% for the lightest squark with a neutralino decay pattern $\tilde \chi^0 \rightarrow \nu_h \snu_{LSP}$, as before. Here, one would have a variety of jet plus multi-lepton final states recoiling against missing transverse energy.

\section{Conclusion}
We have discussed the possibility to observe RH sneutrino DM in the BLSSM, a candidate notably different from the MSSM neutralino. It has been shown before that the sneutrino offers much larger regions of parameter space where the LSP complies with relic density limits in comparison to the MSSM's neutralino. We have explored the ability of ground and space based experiments to detect this candidate. We have found that indirect detection, namely the Fermi-LAT experiment, is beginning to touch the parameter space and a signal may be visible in the next 15 years. We have also seen there are several possible collider signatures, which would be strong hints of this non-minimal SUSY extension. The BLSSM has several features which are distinguishable from the MSSM scenario, and we advocate a more thorough investigation of DM phenomenology in this non-minimal SUSY scenario.

\section*{Acknowledgements}
SM is supported in part through the NExT Institute. The work of LDR has been supported by the STFC/COFUND Rutherford International Fellowship scheme. The work of CM is supported by the `Angelo Della Riccia' foundation and by the Centre of Excellence project No TK133 `Dark Side of the Universe'. The work of SK is partially supported by the STDF project 13858. SK, SJDK and SM acknowledge support from the grant H2020-MSCA-RISE-2014 n. 645722 (NonMinimalHiggs). SJDK and SM acknowledge support from the STFC Consolidated grant ST/L000296/1.  SuK is supported by the `New Frontiers' program of the Austrian Academy of Sciences and by FWF project number V592-N27.

\end{document}